# Review of Damage Models for Environmentally Assisted Cracking on Metals

J.G. Díaz.

1. Introduction

In ordinary fatigue tests about 90 percent of the time is spent in crack nucleation for polished samples and the remaining 10% in crack propagation. Conversely in corrosion fatigue, crack nucleation is accelerated by corrosion; where just 10% of life is just sufficient to form cracks. The remaining time is spent in crack propagation [1]. One can see then, that in corrosion fatigue it is appropriate to evaluate crack propagation behaviour. It was found through a literature search the efforts guided are to perform "break and measure" specimens at the propagation speed to study corrosion fatigue using fracture mechanic's pre-cracked samples[1, 4, 6]. However, new techniques, such as TEM and chemical characterization, have given a new insight to EAC[2].

If a fatigue crack grows in a steady manner under stress concentrator factor critical for fracture, the growth process is termed sub-critical crack[3]. When the maximum applied stress concentration factor surpasses the stress limit for stress corrosion resistance, stress corrosion cracking ads up to the crack growth rate. Usually, small surface defects propagate smoothly under conditions where fracture mechanics does not predict failure. Such is the case of a corrosive environment. Cracks nucleate and propagate below $K_{Ic}$. As a matter of fact, the critical corrosion stress intensity factor, $K_{Iscc}$, may be as low as 1% of $K_{Ic}$.[7, 8]

It can be seen there is a need to include corrosive conditions in damage models that predict accurately, or at least to some reliable extent, the life of a component[2]. This paper reviews some models of fatigue corrosion, SCC, Hydrogen embrittlement, and the unified approach.

2. Review

Literature review found that in the past thirty years, research efforts have been guided to develop empirical models to understand and predict Environmentally Assisted Cracking (EAC), which has four known failures modes [4, 5] : Hydrogen embrittlement, corrosion fatigue (CF), Stress Corrosion, and Liquid metal Corrosion. They all have different models, which aim to predict crack grow, being that they are physical or phenomenological, but lately, some researchers are keen on the unified approach[6, 7] which aims for an analytical solution. However, to date no specific model is completely agreed upon to predict damage under the afore mentioned failure mechanisms. Most of SCC/CF models found in the market fall into these three categories: active path dissolution, film rupture model and hydrogen assisted cracking model[8]. Hydrogen embrittlement modelling aims to understand the dissolution of Hydrogen atoms into metal crystals that produce fragile structures[3,13, 14]. It is agreed it is necessary to know the part's manufacturing processes to improve predictive success in models for SCC due to induced residual stresses[3].

2.1. Models based on active path dissolution and film rupture

Active path dissolution is a process where a strip of material corrodes faster than the rest of the material or structure. Such a strip generally is present where the corrosion resistant alloying elements are separated (mechanically or chemically) mainly from manufacturing process.

### 2.1.1. Ford and Andresen Model

Ford and Andresen[9] stated that the SCC rate depends on the rate of dissolution at the crack tip where the increased tension in the material breaks a thermodynamically stable oxide film. The oxide film forms and ruptures from time to time where such periodicity depends on the strain rate in the core material. The strain is controlled either by creep under constant or by recurrent loading. The bare metal then corrodes along the active path starting a new cycle. The active path dissolution and film rupture model can be used for describing SCC but also for CF. The strain rate at the crack tip also can be owed to material´s built up residual stress during the manufacturing process. The proposed the crack growth rate is based on Arrehnius equation, and it is shown in equation 1:

$$\frac{da}{dN} = \frac{M}{z\rho F} i_o^a ; \qquad \text{eq 1}$$

Where da/dN is average crack grow rate, M is the atomic weight, z is the oxidation number, $\rho$ is the density, F is the Faraday constant, and $i_o^a$ is the base metal dissolution rate parameter.

### 2.1.2. Hall Model

In 2007 Gutman[10] criticised the Ford and Andresen model on the grounds that the oxide´s fracture strain ($\sigma_f$, which is used to demonstrate the model) should be lower than the metal´s elastic limit that is commonly observed for brittle oxide films. In 2009 Hall [11, 12] argued the Ford-Andresen model was unreliable because the crack tip strain rate is dependent of time and tip radius rather than just time. They presented a new model, which included such rate, and it is shown in equation 2.

$$\frac{da}{dN} = \frac{\dot{\varepsilon}_{(r,t)} - \frac{\partial \varepsilon_{ct}}{\partial t}}{|\dot{\varepsilon}_{ct}|} ; \qquad \text{eq 2}$$

Where da/dN is average crack grow rate, and $\varepsilon_{ct}$ is the absolute value crack tip strain rate gradient evaluated at the crack tip radius.

### 2.2. Hydrogen embrittlement based models (H.E.)

Failures associated with H.E. are deferred, include irregular crack growth, often occur without warning, but with severe consequences. The related SCC is usually observed in airless but exposed to aqueous environment components[13].

The mechanism for H.E. is believed to work as follows. Hydrogen atoms can migrate into crystal lattices due to their small size. In those crystals, they go to triaxial tensile stress regions where cracks and notches are likely to appear because tensile stress. Such presence of H atoms in those regions result in increase in local which turn decreases cohesion energy of crystal lattices. A consequence of this is a reduction in ductility, or in other words, an increase in probability of brittle fracture starting at the location of H confinement [13].

In 1991 McEvily[13] summarized several cases of steel failure, including low and high carbon, alloy steel. Back then, it was already recognized that adding to the actual H.E. problem, there was ignorance of the problem by design and maintenance engineers.

In 2001 Hall and Symons[14], published a model for a Ni-Cr-Fe alloy in pressurized water vessels. It was based on crack-growth due by hydrogen assisted creep fracture (HACF) at hydrogen embrittled grain boundaries. The crack growth rate can be calculated as shown in equation 3.

$$\frac{da}{dN} = \frac{r_c \dot{\varepsilon}_{cf\,z}}{\dot{\varepsilon}_{fo}} \sqrt{\frac{C_o}{C_{gb}}} ; \qquad \text{eq 3}$$

Where $r_c$ the radius of fracture zone ahead of crack tip, $\dot{\varepsilon}_{cf\,z}$ the strain rate in creep fracture zone, $\varepsilon_{fo}$ is the fracture strain at a reference H concentration $C_o$ in the grain boundary, and $C_{gb}$ is the grain boundary H concentration.

In 2007 Cheng [15] published a model, that by a Gibbs free-energy balance, and based on Andresen & Ford model, predicted crack growth in neutral pH fluid flow in steel pipelines. The model is presented in equation 4.

$$\frac{da}{dN}_{(\sigma,H)} = k_{\Delta CH} k_H k_\sigma \frac{W}{nF\rho} ; \qquad \text{eq 4}$$

where da/dN is crack grow rate, $k_H$ is Hydrogen´s effect on the anodic dissolution rate in a stress-free environment, $k\sigma$ is stress effect on the anodic dissolution in a H-free environment, $k\Delta_{CH}$ is the effect of Hydrogen concentration difference between stressed and unstressed steel on the anodic dissolution reaction, $\rho$ is density, F is the Faraday constant, W is the molar weight. It is seen the importance given to Hydrogen concentration in this model.

In 2013 Lu [16] published a model for steel pipelines crack growth in near-neutral pH soil environments considering the stress intensity factor, stress ratio, loading frequency, solution pH, and electrochemical potential. In the model, shown in equation 5, the first term represents corrosion contribution and the second the loading effects.

$$\frac{da}{dN} = \frac{Bo}{\sqrt{\ln\left[\frac{C_{cr}^L}{C_B}\right]}} \left\{\frac{\Delta Keq}{f24}\right\}^6 ; \qquad \text{eq 5}$$

Where da/dN is crack grow rate, Bo and $C_{CR}^{L}$ are material constants (related to susceptibility of material to cracking), ΔKeq is equivalent stress intensity factor range and f is loading frequency. Lu claims the model can give decent predictions for working under testing conditions.

## 2.3. Models based on empirical testing

All predictions are done by empirical models which are based on laboratory tests which obviously are limited to testing conditions (materials, pH, temperature, stress, stress variation, etc). Models for SCC and CF are very diverse. There are a few summarized below.

In 1990 a working model was proposed by Garud [17] for alloy 600 tested at 365°C. Shown in equation 6,

$$\frac{da}{dN} = \alpha_o e^{\left[\frac{-Q}{RT}\right]} \dot{\varepsilon}_{net}^{n} \quad ; \quad \text{eq 6}$$

da/dN is crack growth rate, $n = 0.5$, $\acute{\alpha}_0 = 7.338$, $Q = 138.07\ kJ/mol$, $\varepsilon_{cr} = 5 \times 10^{-6}/s$. A few years later, in 1995, Foster et. al. [18] proposed a model that accounted for stress intensity factor in alloy 600 tested at different temperatures. It is shown in equation 7.

$$\frac{da}{dN} = 9{,}216 e^{-\frac{33000}{RT}(Kt-9)^{1.16}} \quad ; \quad \text{eq 7}$$

where R is the Boltzman constant, T is temperature in C and Kt is in MPa√m.

A more developed model was presented the same year by Szklarska and Rebak [19] which included cold work by bending and pH and effects testing alloy 600 at 330 C as shown in equation 8.

$$\frac{da}{dN} = 4{,}76 \cdot 10^{-3} Kt^{1.09} CW^{0.75}; \quad K_t < 30 \text{MPa}\sqrt{m}$$
$$\frac{da}{dN} = 8{,}4 \cdot 10^{-3} Kt^{0.38} pH^{1.67}; \quad K_t > 30 \text{MPa}\sqrt{m} \quad ; \quad \text{eq 8}$$

In 2002 Wei [20] proposed a model to superpose effects of corrosion (da/dN)r and crack growth rate (da/dN)c as shown in equation 9.

$$\frac{da}{dN} = \left(\frac{da}{dN}\right)_c \phi_c + \left(\frac{da}{dN}\right)_r (1-\phi_c); \quad \text{eq 9}$$

where $\phi_c$ is the contributing portion of pure corrosion-fatigue. According to a cross reference search, this model has been used to estimate fatigue crack grown on aluminium, Al–Zn–Mg–Cu alloys, low carbon steel, and austenitic steels.

More recently, in 2013, Yonezu et. al. [21] discussed measures of crack propagation using CPF (corrosion potential fluctuation) and AE (acoustic emission) techniques. In the first one a high-resolution system senses the initiation of localized corrosion of metal in solution, and the non-faradic reactions due to local anodic dissolution are identified with distinctive potential fluctuations, whereas with the second is sensed crack formation right at the moment it begins to propagate. By joining the two techniques, they tried to understand IG-SCC mechanism caused by polythionic acid on stainless steel. They concluded that the stress component plays a principal role during SCC propagation, in addition to the localized dissolution influence on AISI 304 along Cr-depleted grain boundaries.

2.4. Unified Approach

The unified approach recognizes two driving forces for crack growth. ΔK (which creates damage by permanent plastic deformation) and Kmax (which opens and increase crack length). The relation among crack growth rate (da/dN), maximum stress concentration factor (Kmax) and stress concentration range (ΔK) is best described by the unified approach [6] on a 3D plot as shown in figure 1. For a crack to grow, values of Kmax and ΔK cannot reach a threshold value[4] and they are represented by Kmax* and ΔK*.

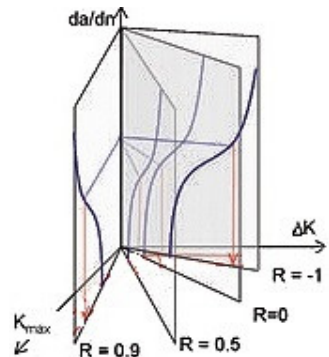

Figure 1. Crack growth rate *da/dN as* a function of ΔK and $K_{max}$

It is expected that the plastic deformation ahead of the tip of the crack leaves a residual stress. When it unloads, it builds up residual compressive stresses at the crack tip preventing it from extra changes. When the load goes to zero, a higher $K_{max}$ must be applied to overcome the residual compressive stresses (and open the crack) so that the measured $\Delta K_{th}$ needed for crack growth is augmented. On the other hand, when the load lowers only a little, the applied load is just plenty to hold the crack open but higher than the residual compressive stress, therefore a change in $\Delta K_{th}$ it is not seen. If some reason the crack cannot close, the same effect is seen. The stress acting on the crack tip is higher than the applied stress. As a result, the effective stress amplitude $\Delta K_{effective}$ is smaller than the applied ΔK. So, a higher $\Delta K_{th}$ will be measured than if the crack faces did not touch each other [6, 4, 8].

Vasudevan [8] classified crack behaviour in corrosive environments in three types as seen in figure 2. The line Kmax*= ΔK* characterises ideal fatigue crack growth; in other words, it represents fatigue crack growth an under inert atmosphere.

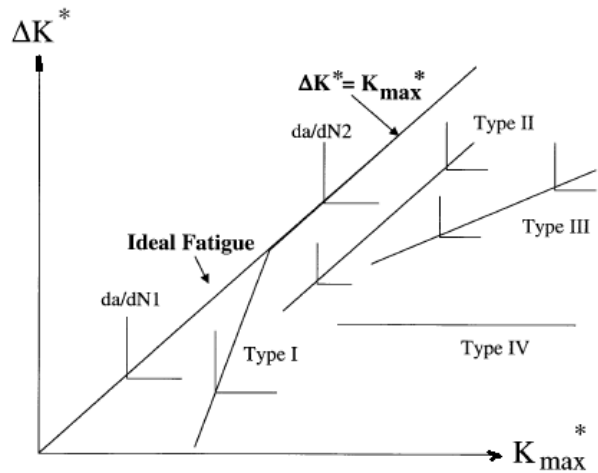

Figure 2. Crack growth trajectory *as* a function of $\Delta K$ and $K_{max}$ (adapted from [8])

In Type I environmental effects are maximum (low crack growth rates but near threshold) and decrease with increasing crack growth rate. Such behaviour is typical of the true corrosion fatigue. Metals surrounded in gaseous environments exhibit four corrosion fatigue steps, which in order are: transference of aggressive media to the crack tip, metal attack the bare crack tip, diffusion of the resulting hydrogen into the metal, and finally metal embrittlement.

The line for Type II shows that the environmental effect is autonomous of crack growth rate, Kmax and/or time because it runs parallel to ideal behaviour (indicated by the K*–Kmax* line). Hence, the mechanism must be different than in Type I. It is believed that the kinetics of Hydrogen transportation, or the reaction at the crack tip, could not be the controlling factor since the result remains constant independent of crack growth rate. Type II behaviour is categorized by an environmental effect that saturates particularly rapidly in relation to the temporary crack advance, and hence delivers a constant influence. Such saturation can be static, or dynamic. It was observed that materials that display Type I behavior at low crack growth rate or low Kmax, change to Type II when those two parameters increase.

In Type III the more $K_{max}$ or crack growth rate increases the bigger the environmental contribution. Such deviation from ideal behaviour, it is associated with environmental effects. Therefore, Type III may be more representative of stress corrosion fatigue process.

At last, Type IV is considered a particular case of Type III where behaviour is controlled completely by Kmax*. This is symptomatic of stress corrosion crack growth rather than stress corrosion fatigue. Crack tip is sharpened by cyclic stress in Type IV accentuating the stress-corrosion effect.

Sadananda[22] stated that the threshold indicates that some minimum stress is necessary to create uncorroded surfaces at a given concentration. That means chemical potential on its own is not enough to create cracks. On the other hand, stress alone can create fresh new cracks. They define a "chemical stress concentration factor'' as the ratio of fracture stress with and without the chemical atmosphere. On another work (Sadananda[23]) used a modified Kitagawa diagram to explain crack growth in term of applied stress. The novelty of this diagram is they separate the stress needed to make a crack advance with the effect of chemical environment as seen in figure 3. It can be seen than for crack sizes are smaller than $a_c$, they cannot grow below the threshold stress (Kth) because since K for such small cracks is less than the Kth* for growth. However, the stress cannot be of mechanical origin only, it could be caused by aggressive media or internal stress or the sum of all three as shown in equation 9. In any case, such stress would produce a K that for a crack to not grow must be smaller than $K_{ISSC}$. The advantage of this diagram is that it describes the behaviour in a smooth coupon while showing both, crack nucleation and crack growth as opposed to traditional precracked-fracture-mechanics coupons where only crack growth is involved.

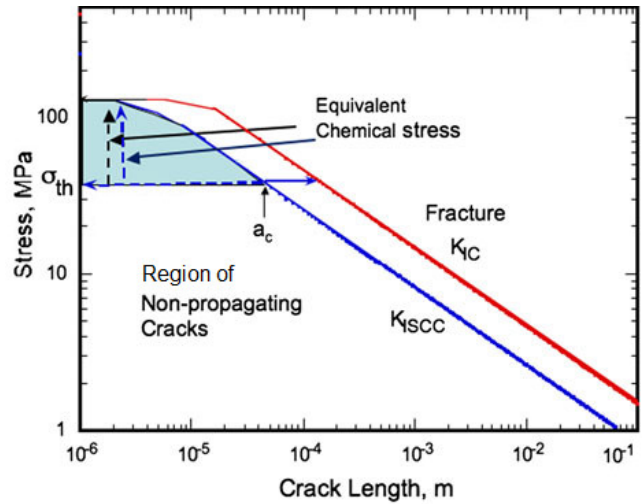

Figure 3. Crack growth vs applied stress (adapted from[23])

Sadananda[23] distinguished three types of curves in a da/dN vs $K_{ISCC}$ diagram. Stage 1 shows a well-defined threshold (a plateau where crack growth rates, stay constant until some K value), Stage 2 shows a slant where crack growth rates are proportional to K, and Stage 3 is a steep slope. Such change in slope could be owed to thickness or H concentration. However, citing other works (Hancock and Johnson, 1966), it is acknowledged that $O_2$ concentration can overtake H effects.

$$K = K_\sigma + K_{int} + K_{chem} \geq K_{ISSC} \text{ ; equation 9}$$

Same authors[23] express the disadvantage of this model, citing to Mikheevskiy[24], is that it depends on an accurate estimation of residual stress produced by previous loading cycles which can possess a challenge. The contribution of internal stress was summarized later by Sanananda[25] using Al 7075-T651 (aged) and SAE 4140 as examples as follows: for a system environment - alloy growth time is inversely proportional to applied stress intensity K. Increasing magnitude of overloads, growth time increases for a constant background K. The internal stress intensity factor can be calculated as a function of the plastic zone difference between the overload (tensile) and the base. Grow time is proportional to the chemical reaction rate at the crack tip or to the diffusion rate ahead of the crack tip.

3. Conclusions

Because crack growth cannot be avoided, engineers need to learn how to live with it. As of this literature review, there is no analytical model agreed upon which performs component´s life prediction. There are multiple models which do crack growth a prediction based on testing conditions. Some materials found to be of interest are for alloy 600 (widely used in power generation and pressure vessel at high temperatures), Al 7075 (high strength-to-density ratio but lower corrosion resistance) and SAE 4140 (multipurpose steel).

Hydrogen embrittelment is recognized as one of the most damaging EAC mechanisms, because its wide presence in different atmospheres. Hence, the great number of models and approaches available for several working conditions.

The requirement of having a close-to-real-life prediction model has driven research to what is called the unified approach. This is driven by the necessity of lowering maintenance / replacing costs by avoiding under-prediction and by lowering the risk of in-service failure by over-predicting a component´s life.

It is agreed that for a crack to grow under a corrosive environment, effects (applied stress, residual stress, and aggressive media) will add up to one damaging force. According to the unified approach, a crack under external stress, internal stress and an aggressive environment will only grow when their sum of the stress intensity factors surpasses the allowed stress intensity factor under corrosive conditions as shown in equation 9.